\documentclass{article}

\usepackage{arxiv}

\usepackage[utf8]{inputenc} 
\usepackage[T1]{fontenc}    
\usepackage{hyperref}       
\usepackage{url}            
\usepackage{booktabs}       
\usepackage{amsfonts}       
\usepackage{nicefrac}       
\usepackage{microtype}      
\usepackage{lipsum}
\usepackage{graphicx}
\usepackage[most]{tcolorbox}
\graphicspath{ {./images/} }

\title{Multimodal Spatial Omics: From Data Acquisition to Computational Integration}

\author{
 Esra Busra Isik\textsuperscript{\textdagger} \\
  Division of Informatics, Imaging \& Data Sciences \\
  School of Health Sciences\\
  University of Manchester\\
  Manchester, United Kingdom \\
  \texttt{esra.isik@postgrad.manchester.ac.uk} \\
   \And
Yusuf Hakan Usta\textsuperscript{\textdagger} \\
  Division of Informatics, Imaging \& Data Sciences\\
  School of Health Sciences\\
  University of Manchester\\
  Manchester, United Kingdom \\
  \And
 Haozhe Liu \\
  Department of Computer Science\\
  University of Manchester\\
  Manchester, United Kingdom \\
    \And
 Maryam Riazi \\
  Division of Informatics, Imaging \& Data Sciences \\
  School of Health Sciences\\
  University of Manchester\\
  Manchester, United Kingdom \\
  \And
William Roach \\
  Department of Microbiology\\
  University Hospitals Tees\\
   Middlesbrough, United Kingdom\\
  \And
   Hongpeng Zhou \\
  Department of Computer Science\\
  University of Manchester\\
  Manchester, United Kingdom \\
    \And
Magnus Rattray\\
  Division of Informatics, Imaging \& Data Sciences\\
  School of Health Sciences\\
  University of Manchester\\
  Manchester, United Kingdom \\
  \And
  Sokratia Georgaka \\
  Division of Informatics, Imaging \& Data Sciences\\
  School of Health Sciences\\
  University of Manchester\\
  Manchester, United Kingdom 
}

\begin{document}
\maketitle

\begin{abstract}
Recent developments in spatial omics technologies have enabled the generation of high dimensional molecular data, such as transcriptomes, proteomes, and epigenomes, within their spatial tissue context, either through cocprofiling on the same slice or through serial tissue sections. These datasets, which are often complemented by images, have given rise to multimodal frameworks that capture both the cellular and architectural complexity of tissues across multiple molecular layers. Integration in such multimodal data poses significant computational challenges due to differences in scale, resolution, and data modality. In this review, we present a comprehensive overview of computational    methods developed to integrate multimodal spatial omics and imaging datasets. We highlight key algorithmic principles underlying these methods, ranging from probabilistic to the latest deep learning approaches.
\end{abstract}
\textsuperscript{\textdagger}\footnote{These authors have contributed equally}


\section{Introduction}\label{sec1}

Spatial omics technologies have fundamentally changed our ability to study biology by enabling molecular profiling of transcriptomes, proteomes, epigenomes, and metabolomes without losing the spatial organization of cells within their native microenvironment \cite{moses2022museum, vandereyken2023methods}. Unlike conventional bulk or single-cell approaches that require tissue dissociation, spatial methods retain the architectural context vital to grasping how cellular phenotypes emerge from local interactions, signaling gradients, and microenvironmental niches \cite{rao2021exploring}. This spatial dimension has proven critical across biomedicine: in oncology, it reveals the heterogeneous organization of tumour microenvironments that influences therapeutic response; in developmental biology, it maps the molecular gradients that pattern tissue formation; in neuroscience, it charts the regional specialization of brain circuits \cite{lewis2021spatial}. Yet no single spatial modality captures the full complexity of a cellular state. Transcriptomics reveals active gene programs but not post-transcriptional regulation; proteomics quantifies functional effectors but misses the regulatory events that produced them; epigenomics exposes chromatin landscapes but not their downstream consequences; metabolomics reflects cellular biochemistry but requires pathway models for interpretation~\cite{vandereyken2023methods, argelaguet2020mofa}. Increasingly, researchers recognize that biological insight demands integration across these molecular layers — linking chromatin accessibility to transcription, protein abundance, and metabolic output within the same spatial context. This recognition has driven the rapid development of both experimental technologies for multimodal spatial profiling and computational frameworks for cross-modal integration.

Integrating multimodal spatial data, however, poses significant computational challenges that distinguish it from conventional unimodal omics analysis. Spatial omics modalities differ fundamentally in resolution, feature dimension and data structure. In addition, when modalities are acquired from serial tissue sections rather than co-profiled on the same slice, spatial correspondence must be computationally inferred through co-registration algorithms that introduce additional uncertainty~\cite{zeira2022alignment}. 

In this Review, we provide a comprehensive overview of multimodal spatial omics integration---from experimental data generation to computational analysis to outstanding challenges. We begin by surveying experimental acquisition strategies, contrasting two fundamental approaches: co-profiling methods that measure multiple modalities from the same tissue section and serial-section workflows that maximize per-modality quality at the cost of requiring computational alignment. We cover sequencing-based and imaging-based platforms across spatial omics modalities, highlighting their respective strengths and limitations. We then introduce key concepts for understanding multimodal data integration, including integration strategies and fusion approaches, before systematically reviewing computational frameworks organized by algorithmic family: probabilistic and statistical inference, matrix factorization and latent variable models, optimal transport and geometric alignment, and deep learning approaches. Throughout, we contextualize methods according to the key computational tasks they address (Box\~1): deconvolution of cell-type mixtures, identification of spatial domains, co-registration of serial sections, prediction of molecular profiles from histology images, and gene imputation. We conclude by identifying fundamental challenges that must be overcome for the field to mature, including standardization of protocols and data formats, development of rigorous benchmarks with ground truth, scalability to terabyte-scale datasets, and interpretable models that connect computational abstractions to testable biological hypotheses, and outline priorities for future development.

\section{Overview of Spatial Multimodal Acquisition Strategies}

Spatial multimodal omics acquisition aims to capture multiple molecular and phenotypic layers from tissue while preserving their spatial organization. Broadly, current strategies fall into two categories: (i) measurements performed on adjacent tissue sections, and (ii) simultaneous measurements on the same tissue section. These strategies differ in experimental design, technical complexity, and the extent to which spatial correspondence between modalities must be reconstructed computationally.

In serial sectioning workflows, consecutive tissue sections are processed using different spatial omics technologies, and the resulting datasets are computationally co-registered to a common coordinate system (Figure 1A). This approach is widely used because it allows each modality to be generated using its optimal protocol. Within this framework, spatial omics platforms can be broadly divided into sequencing-based and imaging-based technologies, each with distinct experimental workflows and data characteristics (Figure 1C).

By contrast, co-profiling strategies perform simultaneous measurements of multiple molecular modalities on the same tissue section, eliminating the need for computational co-registration and preserving exact spatial correspondence between modalities (Figure 1B). Similar to serial sectioning approaches, co-profiling technologies can also be categorized into sequencing-based and imaging-based platforms, reflecting differences in molecular capture, readout mechanisms, and data structure.

\subsection{Measurements on adjacent slices}
\subsubsection{Sequencing-based technologies}
\textbf{Spatial Transcriptomics}

Sequencing-based spatial transcriptomics methods typically use array-based capture slides coupled with next-generation sequencing (NGS) to localize mRNA molecules within the tissue architecture. Spatially resolved transcriptomics was first introduced through Spatial Transcriptomics (ST) technology, which uses a glass slide coated with an array of 1007 spatially barcoded capture probes (‘spots’) with a diameter of 100$\mu m$ containing oligo(d)T primers to bind mRNA molecules from tissue sections~\cite{staahl2016visualization}. This approach was later refined by the 10x Genomics Visium platform, which increased the resolution to 5,000 spots (55$\mu m$ diameter). Recently, the 10x Visium HD platform further advanced spatial transcriptomics to near single-cell resolution, by introducing slides with an increased oligonucleotide barcode density \cite{oliveira2025high}. These technologies are accompanied by high-resolution haematoxylin and eosin (H$\&$E)–stained images acquired from the same tissue section, enabling direct integration of morphological and transcriptomic features. 

Another group of sequencing-based technologies builds on similar core principles but employs DNA-barcoded microparticles (‘beads’) to capture spatially resolved gene expression profiles. Slide-seq introduced a strategy where 10$\mu m$ DNA-barcoded beads are randomly deposited on a rubber-coated glass slide to form a dense array termed as ‘puck’~\cite{rodriques2019slide}. While Slide-seq offers a highly resolved spatial transcriptomics method, its low transcript detection sensitivity limits application across a wide range of biological problems. Slide-seq V2 addresses this limitation by offering an order of magnitude higher sensitivity through enhancements in the barcoded bead synthesis, array indexing and library preparation, enabling wider application of this technology~\cite{stickels2021highly}. High-Definition Spatial Transcriptomics (HDST) relies on a similar strategy, using a randomly ordered bead array, where barcoded poly(d)T oligonucleotides deposited into 2$\mu m$ wells~\cite{vickovic2019high}. Unlike Slide-seq technologies, HDST provides matched H$\&$E images. 

\begin{figure*}[ht]%
\centering
\includegraphics[width=1\linewidth]{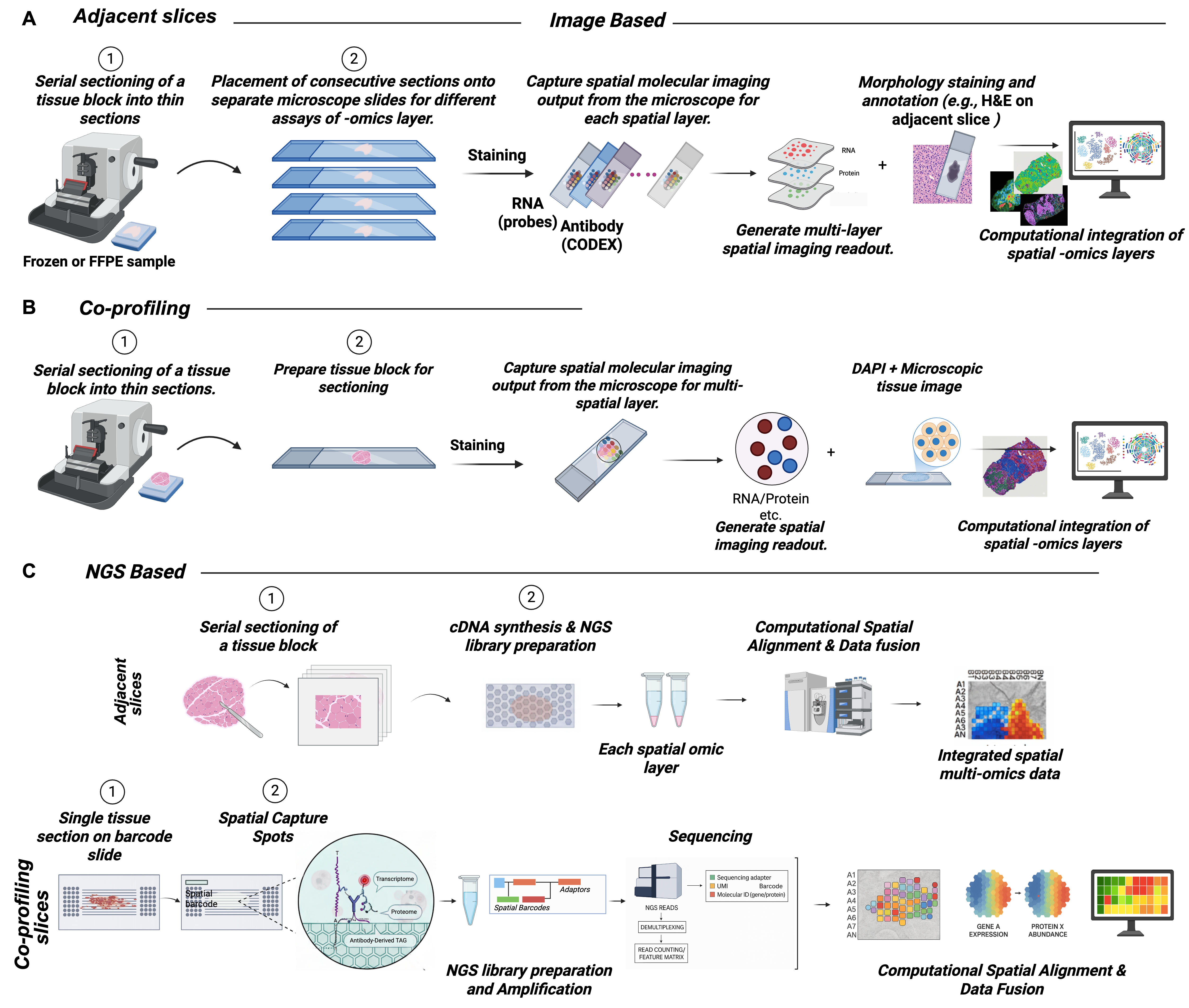}
\caption{\textbf{Experimental workflows for spatial multi-omics data acquisition.} The figure illustrates three major experimental approaches for generating multimodal spatial omics data. \textbf{A:} Serial sectioning workflow where consecutive tissue sections are processed using different modalities. Following tissue sectioning, individual slices undergo staining and are analyzed using distinct spatial omics technologies (e.g., spatial transcriptomics on one section, spatial proteomics on another). Subsequent computational co-registration aligns the datasets to a common coordinate system, enabling cross-modality integration of spatially resolved molecular profiles. \textbf{B:} Measuring several molecular types at once on the same piece of tissue. Once the tissue has been made ready and stained, both RNA and protein (or other types) are taken from the same places in space, which removes the need for computer-based matching and keeps exact location links between measurements. \textbf{C:} Comparison of sequencing-based and imaging-based spatial omics platforms. The NGS-based workflow (left) shows tissue sectioning, spatial barcoding via array or bead-based capture, cDNA synthesis and library preparation, followed by next-generation sequencing and computational analysis to generate integrated spatial multi-omics data. The image-based workflow (right) depicts in situ detection methods using sequential rounds of hybridization or antibody labeling, high-resolution imaging, and image analysis to produce spatially resolved molecular profiles with subcellular resolution.}
\label{}	
\end{figure*}

\textbf{Spatial Epigenomics}

The spatial organization of the epigenome – DNA regulatory elements and chromatin modifications – is crucial for understanding gene regulation in development and disease. Only recently have methods emerged to profile epigenetic marks with spatial resolution~\cite{zhang2024mass}. Pioneering techniques are adapting assays like ATAC-seq (for chromatin accessibility) and CUT$\&$Tag (for histone modifications or transcription factor binding) to tissue sections~\cite{farzad2024spatially}.

In the standard ATAC-seq, transposase enzymes cut open chromatin regions and insert sequencing adapters to profile open chromatin. Spatial ATAC-seq introduces these transposases into fixed tissue sections and then uses spatial barcoding (often with a DBiT-seq style microfluidic grid) to index the DNA fragments by location. In this approach, transposase (Tn5) flows across the tissue to fragment DNA in situ, then two sets of barcodes (A and B channels) are applied to the tissue as described for DBiT, so that each cut fragment gets tagged with a location-specific barcode. Sequencing these fragments yields a map of open chromatin regions across the tissue. Spatial ATAC-seq has achieved $\sim$20$\mu$m resolution (roughly single-cell level) in tissue sections.

Cleavage Under Targets and Tagmentation (CUT$\&$Tag ) is an assay where an antibody targeting a specific histone modification directs a transposase to those genomic sites, which are then cut and tagged for sequencing. Spatial CUT$\&$Tag adapts this by performing the antibody binding in tissue sections, then using microfluidic barcoding to assign spatial addresses. The result is a map of a particular histone mark’s enrichment across the tissue. A breakthrough study applied spatial CUT$\&$Tag to mouse embryos for three key marks (H3K4me3 active promoters, H3K27ac active enhancers, H3K27me3 repressive heterochromatin) at $\sim$ 20$\mu m$ resolution. It revealed, for instance, that developing brain regions had distinct epigenetic landscapes: some neuronal layers had strong H3K27ac at certain enhancers correlating with neuronal subtype specification, whereas other regions showed H3K27me3 silencing developmental genes in cells that had terminally differentiated. In another application, spatial CUT$\&$Tag on diseased tissue could highlight fibrotic regions with abnormal chromatin states, helping explain pathological gene expression. A key challenge of these methods is data sparsity. Even single-cell ATAC-seq exhibits considerable sparsity, which can be further exacerbated in tissue contexts. Aggregating data across neighboring cells within a local region can help mitigate this limitation ~\cite{deng2022spatial2,liu2024spatial,cao2024systematic}.


\subsubsection{Imaging-based technologies}
\textbf{Spatial Transcriptomics}

In-situ hybridization (ISH) techniques enable the quantification, localization and visualization of individual RNA molecules using nucleic acid probes that are complementary to the target RNA sequence~\cite{jensen2014technical}. Early ISH methods employ probes labeled with radioactive isotopes, such as tritium. These were later replaced by fluorescently labeled probes leading to improved resolution, enhanced safety and easier imaging. This advancement gave rise to Fluorescent in Situ Hybridization (FISH), and subsequently to single-molecule FISH (smFISH), which allows detection and counting of individual RNA transcripts within cells. Since each gene requires a unique fluorophore, this approach is limited to a low number of genes (typically 3-5). Sequential FISH (seqFISH) and its improved version, seqFISH+, are extensions of smFISH that enable the detection of thousands of genes by repeatedly reusing a set of fluorophores across multiple rounds of hybridization~\cite{shah2017seqfish,eng2019transcriptome}. 

An alternative method that overcomes the limitations of the smFISH approach is Multiplexed Error-Robust FISH (MERFISH). Instead of assigning each gene with a unique fluorophore or reusing the same fluorophores across multiple cycles, MERFISH uses combinatorial labeling and sequential imaging where error-robust binary barcodes are assigned to individual RNA species which are subsequently read through sequential hybridization rounds~\cite{chen2015spatially}. Key commercially available FISH-based platforms include Vizgen’s Merscope (MERFISH) and NanoString CosMx (smFISH).  

In situ sequencing (ISS) is another imaging-based technology that enables the direct sequencing of RNA molecules within tissue. ISS relies on barcoded padlock probes that hybridize to specific RNA targets, followed by ligation and amplification through rolling circle amplification, generating rolling circle products that can be visualized via imaging-based sequencing methods~\cite{lee2022direct}. Xenium, a commercially available ISS-based platform provided by 10x Genomics, provides subcellular resolution for panels of up to 5,000 genes~\cite{janesick2023high}. 

\textbf{Spatial Epigenomics}

In spatial epigenomics, chromatin tracing uses sequential DNA FISH to trace the three-dimensional paths of chromosomes in nuclei, providing a spatial map of genome folding (a technique sometimes called “Hi-M” for in situ Hi-C visualization). These have shown how certain loci come together in “hubs” inside the nucleus of specific cell types. This imaging technique achieves extremely high-resolution (30 nm between probes) and has been used to map the 3D genome in single cells of tissue, although its throughput remains limited~\cite{su2020genome}.

\textbf{Spatial Proteomics}

Imaging-based spatial proteomics comprises a spectrum of technologies that visualize and quantify proteins directly within tissues. These methods differ in how proteins are tagged or detected, the imaging modality used, and the achievable resolution and multiplexing. They can broadly be grouped into antibody-based approaches, metal-tagged immunohistochemistry coupled to mass spectrometry, and label-free workflows using LC-MS/MS or imaging mass spectrometry (IMS).

Multiplexed fluorescence imaging remains a widely used strategy, relying on repeated cycles of antibody staining and imaging to overcome the limited spectral range of fluorophores. In each cycle, antibodies conjugated to fluorophores are applied, imaged, and then chemically quenched or removed before the next set. Techniques such as Array Tomography~\cite{Micheva2007}, MELC~\cite{Schubert2006}, MxIF~\cite{Gerdes2013}, CycIF~\cite{Lin2015}, t-CyCIF~\cite{Lin2018,Du2019}, 4i~\cite{Gut2018}, and IBEX~\cite{Radtke2020,Radtke2022} have been applied across diverse tissues and are particularly valuable for mapping the tumour immune microenvironment. While they achieve subcellular resolution, very high-plex experiments require many cycles, increasing processing time and risking antigen loss, incomplete quenching, and antibody cross-reactivity. The large dynamic range of protein abundance also challenges fluorescence sensitivity~\cite{Milo2010}.

DNA-barcoded antibody methods improve scalability by separating antigen recognition from signal readout. After a single staining step, complementary fluorophore-labeled oligonucleotides are hybridized and removed in rapid cycles, as in Exchange-PAINT~\cite{Jungmann2014}, DEI~\cite{Wang2017}, and CODEX~\cite{Goltsev2018,Schurch2020}. Signal amplification strategies, such as Immuno-SABER~\cite{Saka2019} and isHCR~\cite{Lin2018HCR}, increase sensitivity to low-abundance proteins, while metal isotope readouts, as in SABER-IMC~\cite{Hosogane2023}, combine the advantages of DNA barcoding with imaging mass cytometry.

Attaching ionizable metal isotopes to antibodies enables highly multiplexed “next-generation IHC” for mass spectrometry imaging~\cite{Rimm2014}. In multiplexed ion beam imaging (MIBI), a primary ion beam releases secondary ions from metal-labeled antibodies for magnetic sector or time-of-flight analysis, achieving up to $\sim$100-plex dynamic range~\cite{Angelo2014} and revealing tumour–immune architecture at subcellular resolution~\cite{Keren2018,Keren2019}. High-definition MIBI further improves resolution to $\sim$30\,nm~\cite{RoviraClave2021}. Imaging mass cytometry (IMC) instead uses laser ablation to deliver tags into a TOF analyser, analogous to CyTOF single-cell assays~\cite{Bendall2011}. It enables simultaneous imaging of dozens of proteins and post-translational modifications~\cite{Giesen2014}, and integration with RNAscope reveals mRNA–protein concordance in tumours~\cite{Schulz2018}. These methods offer high multiplexing with minimal background but require vacuum-compatible samples, lack intrinsic signal amplification, and involve destructive acquisition.

Label-free LC-MS/MS spatial proteomics combines sensitive mass spectrometry with precise tissue sampling, often via laser capture microdissection (LCM)~\cite{Ahmad2023,Tan2024,Budnik2018,Zhu2018,Li2018,Schoof2021}. Pixel-by-pixel strategies divide tissues into voxels for individual processing, quantifying hundreds to thousands of proteins per voxel~\cite{Petyuk2007,Piehowski2020,Ma2022,Davis2023}. Region-of-interest approaches focus on histologically defined areas, as in LCM-SISPROT, which profiles distinct cell types from very small regions~\cite{Xu2018,Liang2018}.  Deep Visual Proteomics (DVP) integrates high-resolution imaging, AI-based cell classification, and ultrasensitive MS to map thousands of proteins at single-cell resolution [64], while scDVP extends this to single-cell spatial maps \cite{Mund2022}. These approaches enable generation of comprehensive spatial protein atlases\cite{Rosenberger2023}. 

IMS-based proteomics offers an untargeted, label-free view of protein distributions by directly acquiring mass spectra at defined spatial coordinates~\cite{burnum2008,han2019}. Top-down IMS analyses intact proteoforms using MALDI, nanoDESI, or LESA~\cite{addie2015,buchberger2018,griffiths2018}, with recent advances enabling near-cellular resolution in human tissues~\cite{zemaitis2022,chen2021,liu2021}. Bottom-up IMS digests proteins in situ and images the resulting peptides~\cite{diehl2015}, a particularly useful approach for FFPE tissues~\cite{stillger2024}. Workflow optimizations and enzyme combinations expand coverage~\cite{heijs2015}, and LC-MS/MS of adjacent sections aids peptide identification~\cite{maier2013,huber2018,li2024}.

\textbf{Spatial Metabolomics}

Metabolites and small molecules present another frontier for spatial omics. While not explicitly mentioned alongside transcriptomics and proteomics as frequently, spatial metabolomics refers to mapping the distribution of metabolites (like lipids, sugars, signaling molecules, drug compounds) within tissues. The primary technologies for this are in the realm of mass spectrometry imaging, as methods analogous to FISH do not exist for small molecules (they are too varied and generally can’t be tagged in situ without perturbation). 
The workhorse is MALDI Imaging Mass Spectrometry, already described above, which can simultaneously detect a broad array of metabolites in tissue without labels~\cite{lee2025spatial}. Another technique is DESI (Desorption Electrospray Ionization) Imaging, which uses a stream of charged solvent droplets to desorb molecules from the tissue surface for MS analysis. DESI can profile lipids and metabolites directly from tissue at $\sim$50–200$\mu m$ resolution and has the advantage of minimal sample prep (no matrix coating needed)~\cite{bourceau2023visualization,zhu2022advances}.
Because metabolites are so chemically diverse, no single imaging method captures all – but MS-based approaches together provide a powerful window ~\cite{zhang2024mass}. The data analysis often involves overlaying metabolite maps with anatomical features to interpret, for example, how a drug distributes in a tumour or which metabolic pathways are active in a microenvironment. For computational biologists, spatial metabolomics raises interesting challenges in spectral deconvolution and metabolite identification directly from tissues, as well as integration with transcriptomic or proteomic maps (to link enzymes with their products spatially). Although still a specialized field, spatial metabolomics is becoming important in pharmacology (drug localization and effect mapping) and pathology (metabolic reprogramming in disease contexts).

\subsection{Simultaneous Measurements on the same section}

\subsubsection{Spatial Transcriptomics and Spatial Proteomics}

Current methods for co-detecting spatial transcriptomics and proteomics on the same tissue section build on principles from single-modality approaches but are often limited by low spatial resolution or a restricted number of detectable proteins. The standard 10x Genomics Visium platform supports whole transcriptome profiling alongside immunofluorescence-based detection of one or two protein markers. The Visium CytAssist workflow expands this by using spatially barcoded probes for mRNA detection in combination with oligo-conjugated antibodies for protein profiling. While it enables transcriptome-wide coverage, protein detection is limited to a pre-validated panel of 35 markers, with the option to include additional custom antibodies, which require careful optimization. Spatial PrOtein and Transcriptome Sequencing (SPOTS) relies on Visium’s poly(A) capture technology in combination with DNA-barcoded antibodies to enable simultaneous spatial profiling of the whole transcriptome alongside a panel of over 30 protein markers~\cite{ben2023integration}. Similarly, Spatial Multi-Omics (SM-Omics) builds on the original Spatial Transcriptomics platform for transcriptome-wide measurement although at a lower spatial resolution (100$\mu m$) while it incorporates fluorescently labeled or DNA-barcoded antibodies for detection of 6 proteins~\cite{vickovic2022sm}. Spatial-CITE-seq enables simultaneous spatial mapping of the whole transcriptome and $\sim$ 200–300 proteins at 25$\mu m$ pixel resolution by staining tissue sections with poly(A)-tailed antibody-derived tags (ADTs), followed by in situ barcoding of both ADTs and mRNAs~\cite{liu2023high}.

Imaging-based methods have been extended to enable spatial co-profiling of transcriptomes and proteins at subcellular resolution, though typically offer lower gene and protein throughput compared to sequencing-based approaches. 10x Genomics Xenium supports spatial co-profiling of RNA and protein using a pre-defined panel of up to 480 genes and 27 proteins focused on the human immune system and tumour microenvironment. Similarly, Vizgen’s MERSCOPE allows simultaneous spatial profiling of hundreds to thousands of RNA transcripts alongside up to 6 proteins through oligo-conjugated secondary antibodies. Spatial Molecular Imaging (SMI), based on CosMx (a FISH-based method), enables in situ detection of up to 1,000 genes and 64 proteins at subcellular resolution by hybridizing fluorescent molecular barcodes across multiple hybridization cycles~\cite{he2022high}.

\begin{figure*}[ht]%
\centering
\includegraphics[width=1\linewidth]{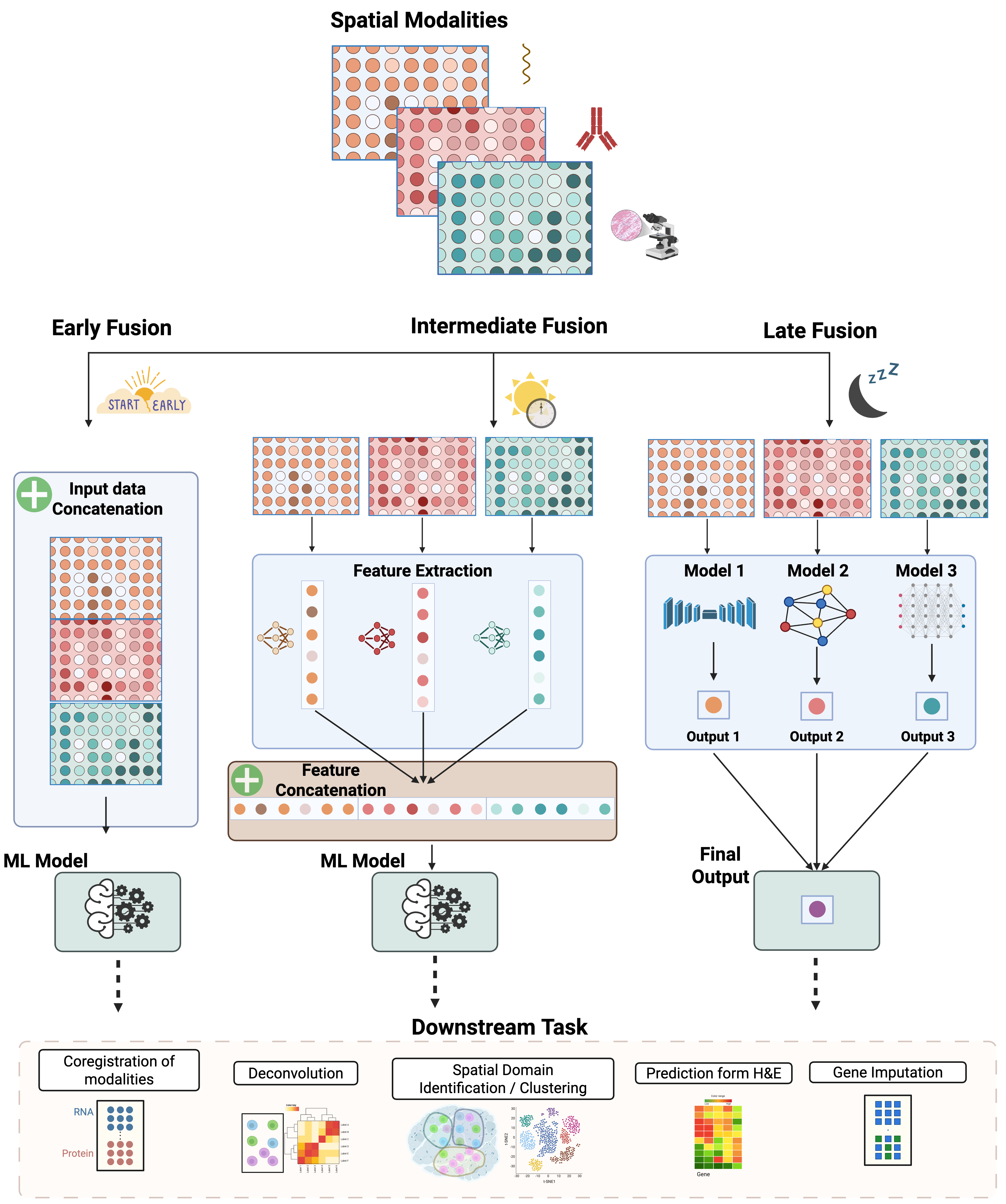}
\caption{ \textbf{Fusion strategies for multimodal spatial omics integration.} Three principal approaches for combining information from multiple spatial omics modalities. \textbf{Early Fusion:} Raw features from different modalities (e.g., spatial transcriptomics, proteomics, histology) are concatenated into a single input matrix before being processed by a unified machine learning model. \textbf{Intermediate Fusion:} Each modality is first processed independently to extract modality-specific latent representations, which are then concatenated at the feature level and fed into a downstream model. This approach allows modality-specific preprocessing while enabling the model to learn cross-modal interactions. \textbf{Late Fusion:} Separate models process each modality, returning independent predictions that are later aggregated (e.g., via voting, averaging, or weighted combination) to generate the final output. The bottom panel illustrates key downstream computational tasks enabled by these integration strategies: co-registration of modalities, cell-type deconvolution, spatial domain identification and clustering, prediction of molecular profiles from H\&E images, and generative modeling for data imputation and augmentation.}
\label{}	
\end{figure*}

\subsubsection{Spatial transcriptomics and spatial epigenomics}
Co-profiling of spatial gene expression and histone modifications or chromatin accessibility has been made possible through technologies motivated by DBiT-seq. MISAR-seq simultaneously profiles spatial gene expression and chromatin accessibility by combining principles from DBiT-seq and SHARE-seq~\cite{ma2020chromatin}, using a 2,500 microfluidic grid (50$\mu m$ pixel resolution) design where channel specific DNA barcodes are flowed through two perpendicular channels to spatially index Tn5-labeled open chromatin regions and reverse-transcribed mRNAs~\cite{jiang2023simultaneous}. The technology has been demonstrated on the fetal mouse brain to study cell fate determination and spatiotemporal regulatory dynamics during development, however, according to the authors, MISAR-seq is compatible with a wide range of tissue types. 
Similarly, to further investigate the epigenetic mechanisms underlying spatially resolved gene expression regulation, genome-wide co-mapping of the epigenome and transcriptome has been developed through spatial ATAC–RNA-seq  at near single-cell resolution (20$\mu m$ pixel microfluidic channel array chip). These methods enable co-profiling of chromatin accessibility or histone modifications (e.g., H3K27me3, H3K27ac, H3K4me3) alongside mRNA expression at cellular resolution via deterministic co-barcoding. By integrating the chemistry of spatial ATAC-seq or CUT\&Tag with spatial transcriptomics, these approaches provide comprehensive insights into the spatial coordination of epigenetic regulation and transcription. These technologies have been demonstrated in embryonic and juvenile mouse brains and adult human hippocampus to dissect the dynamic interplay between chromatin states and gene expression during development~\cite{li2025spatially,zhang2023spatial}.

\section{Overview of Multimodal Spatial Omics Data Integration Strategies}

There are a wide range of algorithms and strategies to integrate spatial omics modalities, and the choice of approach often depends on the specific acquisition strategy and the downstream computational task (Box 1). These methods can be conceptualized along two axes: the integration strategy and the fusion strategy. According to~\cite{argelaguet2021computational} and based on the experimental design, integration can be classified as horizontal, vertical and diagonal. In multimodal spatial omics, horizontal integration refers to measuring the same modality across different donors, replicates, or different experimental conditions. In this case, the common dimension \textrm{--} or the anchor \textrm{--} is the shared feature and alignment is performed between samples to enable comparative analysis across different conditions or samples. 
Vertical integration occurs when different spatial modalities (e.g. spatial transcriptomics and spatial epigenomics) are measured either on the same tissue slice through spatial co-profiling or on serial slices with subsequent co-registration (box 1). In this scenario, the anchor is the spatial spot, allowing cross-modal analysis across tissue and spatial niches. Diagonal integration is used when there is no anchor between the different datasets, for example when integrating spatial transcriptomics from one tissue sample with spatial proteomics from the same tissue type but from different donor. In this case, a common computational strategy is to infer a shared latent space which captures correspondences between the different spatial layers. 

The second axis concerns the fusion strategy (Figure 2). Whereas the integration strategy defines the conceptual framework for aligning multiple modalities along a common dimension, the fusion strategy describes when and how information from each modality is computationally combined. In early fusion, the modality-specific features are concatenated into a single matrix before being fed into the final integration model. Early fusion, also known as feature-level fusion, is simple and straightforward, as a single model is used to learn a joint representation of the multimodal data. Moreover, in deep learning approaches, its lower architectural complexity makes it a popular and convenient choice. However, early fusion also introduces important limitations. Spatial multimodal omics datasets are inherently heterogeneous and span different scales, distinct noise characteristics and statistical distributions and data structures (e.g. histology image versus counts-based spatial transcriptomics). Effective modality-specific preprocessing and normalization are therefore essential and can strongly influence downstream model performance. Moreover, by simply concatenating all modalities into a single input, the model loses information about modality identity, hindering its ability to infer complex or nonlinear cross-modality or inter-modality interactions.   

In late fusion, or decision-level fusion, each modality is processed independently through its own model, and integration occurs only at the prediction level, where the individual model outputs are combined into a final decision. This approach closely resembles ensemble methods. Common aggregation techniques include majority voting, averaging probabilities, or weighted average. The modularity of the late fusion approach offers a flexible way of integrating multimodal spatial omics data, allowing different model architectures \textrm{--}and therefore diverse data types\textrm{--}, to be used for each modality. Furthermore, new modalities can easily be added in the pipeline without requiring modification or retraining of the existing modality-specific models. Missing modalities is a phenomenon commonly observed in spatial multimodal omics; for example, in integrating spatial transcriptomics and spatial proteomics from serial slices, some samples may lack one of the modalities. Late fusion can handle such scenarios because the predictions are generated independently by each model. Nevertheless, unless there is a cross-modality independence assumption, this "first model, then integrate" strategy can overlook cross-modality interactions which are often essential for downstream analysis and biological interpretation (Figure 2).  

Intermediate fusion, also called representation-level fusion, is a hybrid of early and late fusion that aims to learn a joint representation of the different spatial omics modalities. In this approach, each modality is first processed independently to extract modality-specific latent representations, which are then combined into a joint representation matrix that is fed into the final model for downstream tasks.  Methods for merging the individual representations include simple concatenation, element-wise operations (summation, averaging or multiplication) as well as more advanced approaches based on attention mechanisms. Modality specific preprocessing allows the use of tailored models for each modality, facilitating the extraction of high-quality latent features that capture within modality correlations. These features, in turn, enable the final model to learn more complex and nonlinear interactions both within and across modalities. A key limitation of intermediate fusion is modality imbalance, where features from one modality might dominate the joint representation, leading to biased embeddings. Approaches such as modality-specific weighting can mitigate these effects and help the model to perform more balanced predictions (Figure 2).

\section{Computational Frameworks for Multimodal Spatial Omics \mbox{Integration}}
The integration strategies outlined above provide a conceptual framework for how multimodal spatial omics and imaging datasets can be aligned across experimental designs, resolutions, and feature spaces. However, translating these strategies into practical analyses requires concrete computational implementations. Figure 3 provides an overview of the rapidly expanding ecosystem of computational tools, summarizing the modality combinations they integrate and the downstream computational tasks, as defined in box 1.

In the following subsections, we review the major classes of computational frameworks for multimodal spatial omics integration, organizing methods by their underlying algorithmic principles while highlighting how each method addresses key computational tasks in spatial omics analysis (Box 1 and Figure 3). 

\begin{tcolorbox}[
  colback=gray!12,
  colframe=gray!12,
  width=\linewidth,
  boxrule=0pt,
  arc=0pt,
  left=6pt,
  right=6pt,
  top=6pt,
  bottom=6pt,
  breakable
            ]

\textbf{Box 1 | Key Computational Tasks}

\vspace{0.6em}

\textbf{Co-registration of modalities}\\
Spatial omics datasets obtained from serial tissue sections must be co-registered to a common coordinate system before integration, with histology-stained images often serving as proxies for this co-registration.

\textbf{Deconvolution}\\
Some of the sequence-based spatial transcriptomics technologies capture mRNA at multicellular rather than single-cell resolution (for example, 10X Visium, Slide-seq). As a result, each spot can contain a mixture of cell types. Computational deconvolution methods have therefore been developed that infer the cell-type composition at each spatial location. Most of these approaches can be viewed as multimodal: some rely on annotated single-cell RNA sequencing (scRNA-seq) reference that link cell-type specific transcriptomic profiles to spatial spots, while others integrate histological image features extracted from paired H$\&$E-stained images, to improve deconvolution performance.

\vspace{0.5em}
\textbf{Spatial Domain Identification}\\
In spatial omics, spatial domains are contiguous tissue regions where cells share similar molecular expression patterns and biological functions. These domains are biologically important because they help identify distinct cell states, disease-associated regions or microenvironments, as well as distinct functional areas such as fibrotic or hypoxic niches. Identifying in the spatial domain through the integration of various modalities in spatial omics, like transcriptomics, proteomics, and imaging, provides greater insight into how tissues are organized and function.

\vspace{0.5em}
\textbf{Prediction of omics from histology images}\\
Histology imaging is usually routinely performed in clinical practice and serves as cornerstone of medical diagnosis. In contrast, spatial omics assays remain costly and are rarely implemented in clinical settings. With the growing availability of paired histology and spatial omics datasets, deep learning methods have emerged that can predict spatially resolved transcriptomic and proteomic profiles directly from H$\&$E-stained images. Most of these approaches adopt a tile-based representation of the H$\&$E image, whereby whole-slide images are segmented into fixed-size patches that are spatially aligned to the resolution of the underlying spatial omics technology, most commonly matching the diameter of a 10x Visium spot. These image patches are then used as inputs to deep learning models to learn mappings between morphological and  molecular features. These approaches aim to leverage histology as a low-cost proxy for expensive spatial omics measurements. 
\vspace{0.5em}

\textbf{Gene Imputation $\&$ Cross-Modality Translation}\\
In situ hybridization–based spatial transcriptomics techniques provide single-cell–resolved spatial gene expression, but they are limited by low gene throughput. Gene imputation methods address this limitation by leveraging scRNA-seq reference datasets combined with spatial transcriptomics to infer spatial gene expression of genes that are not directly measured. 
\\
Cross-modality translation methods typically rely on generative models trained on paired spatial multimodal datasets to learn correspondences between spatial modalities and predict one modality from another. These approaches enable multimodal spatial omics integration in settings where one or more modalities are missing.

\end{tcolorbox}
\vspace{1em}

\begin{figure}
    \centering
    \includegraphics[width=1.15\linewidth]{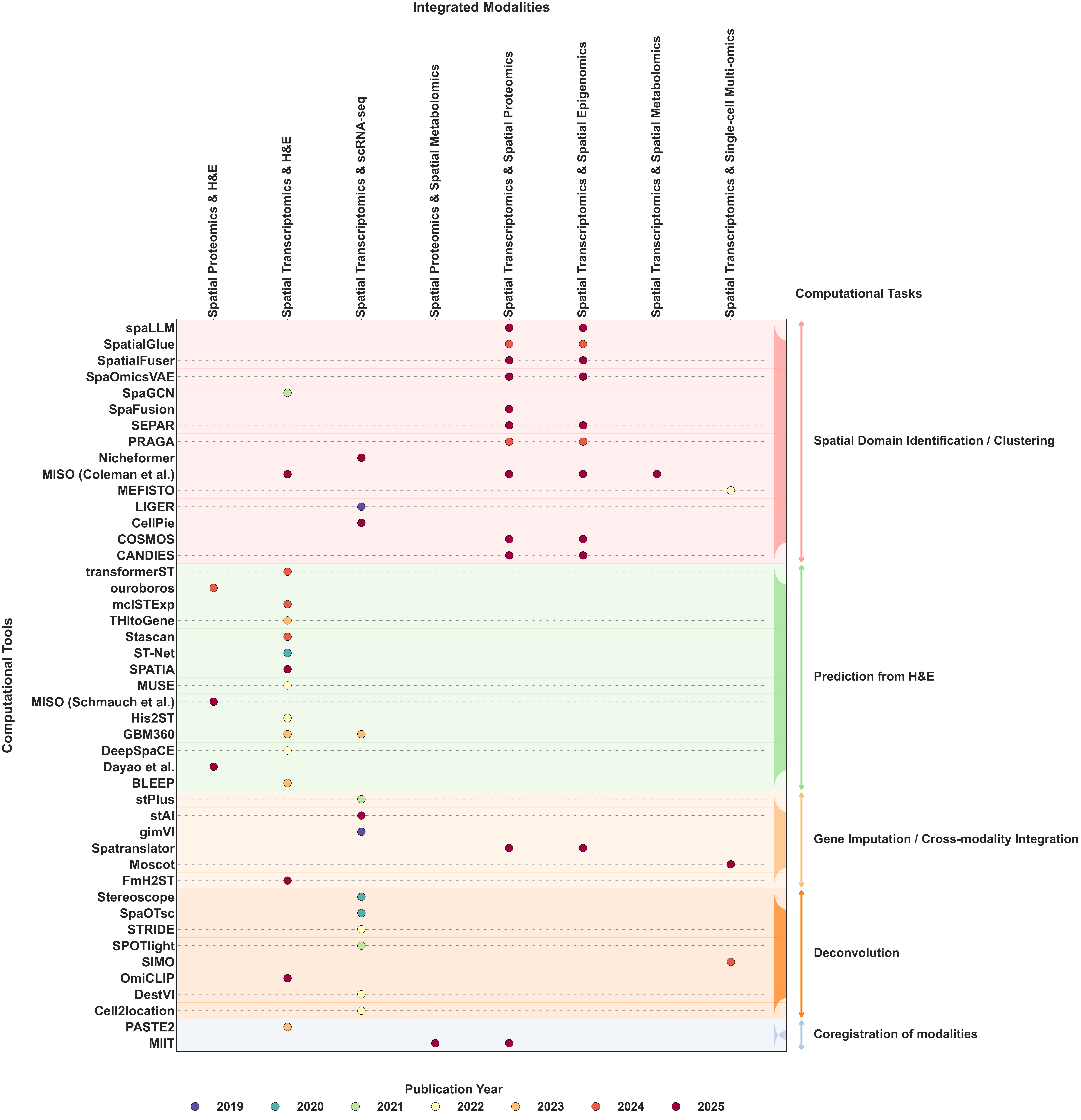}
    \caption{\textbf{Landscape of computational tools for multimodal spatial omics integration.} Overview of published computational methods mapped by the modality combinations they integrate and the downstream analytical tasks they address. Methods are categorized by five key computational tasks: co-registration of modalities, cell-type deconvolution, spatial domain identification, prediction of molecular profiles from H\&E images, and generative modeling. The timeline illustrates the rapid expansion of the field, with a marked acceleration in method development from 2022 onwards. Notably, the integration of spatial transcriptomics with H\&E images and scRNA-seq references represents the most mature area, with numerous established tools available. In contrast, computational strategies for integrating of spatial epigenomics, spatial metabolomics, and multi-modal spatial proteomics remain comparatively underexplored, highlighting key opportunities for future methodological development. Note: some tools support multiple computational tasks; for ordering and row shading we show the primary task.
    }
    \label{fig:placeholder}
\end{figure}

\subsection{Probabilistic and Statistical Inference Methods}

Multimodal spatial omics integration can be formulated as a probabilistic inference problem, in which measurements from multiple data modalities are assumed to arise from shared latent variables that capture underlying biological structure \cite{Kirk2012, Argelaguet2018}. In this setting, a latent variable \(z_n\) is associated with each spatial location or spot \(n\) and represents shared biological properties such as cell-type composition, cell states, or low-dimensional transcriptional programs (Figure 4A). Each modality \(m \in \{1,\ldots,M\}\) is described by a modality-specific likelihood that captures its distinct measurement process, noise characteristics, and data distribution. Let \(\{X^{(m)}\}_{m=1}^M\) denote multimodal spatial measurements (e.g.\ spatial transcriptomics gene expression counts and spatial proteomics protein intensities). The generative model is defined as
\begin{equation}
X^{(m)}_{n} \sim p\!\left(X^{(m)}_{n} \mid z_{n}, \Theta^{(m)}\right),
\quad
z_{n} \sim p(z_{n} \mid \Phi),
\end{equation}
where \(z_n\) represents latent variables encoding the underlying biological state at spatial location \(n\) (e.g.\ cell-type composition or latent factors). The parameters \(\Theta^{(m)}\) denote modality-specific generative parameters, such as latent factor loadings linking \(z_n\) to observed gene or protein measurements, as well as additional parameters governing the observation model (e.g.\ dispersion or noise parameters). These parameters may be learned directly from the data or informed by external reference profiles, such as cell-type--specific expression or protein abundance signatures. The prior over latent variables is specified by \(\Phi\).

The multimodal integration task then corresponds to estimating the posterior distribution
\begin{equation}
p\!\left(z_n \mid X^{(1)}_n, \ldots, X^{(M)}_n \right),
\end{equation}
which is generally intractable and therefore approximated using variational inference or related Bayesian approximation techniques.

Several spatial deconvolution tools adopt this probabilistic formulation to integrate spatial transcriptomics with single-cell RNA-seq references. Stereoscope \cite{Andersson2020} models spatial gene expression as a mixture of reference-derived cell-type profiles under a Negative Binomial likelihood, enabling estimation of cell-type proportions together with uncertainty. Cell2location \cite{Kleshchevnikov2022cell2location} extends this approach using a hierarchical Bayesian model that explicitly accounts for technical effects and overdispersion, improving robustness and sensitivity, particularly for rare cell types. DestVI further generalizes probabilistic deconvolution by introducing continuous latent variables to model within–cell-type state variation, combining Bayesian inference with neural parameterization in a variational framework \cite{Lopez2022destvi}.

Hybrid approaches combine probabilistic inference with deep learning for spatial prediction tasks. MUSE integrates histological images (H\&E) with spatial transcriptomics using a framework that includes probabilistic components alongside neural networks, enabling prediction of gene expression from tissue morphology while retaining a distributional view of the data \cite{bao2022integrative}. Similarly, CANDIES incorporates probabilistic inference within a broader deep learning–based multi-omics integration pipeline, allowing uncertainty-aware spatial domain identification while integrating spatial transcriptomics with spatial epigenomic or proteomic modalities \cite{Liu2025-mz}.

More broadly, these spatial probabilistic models are conceptually related to classical Bayesian multi-view approaches such as Multiple Dataset Integration (MDI) \cite{Kirk2012} and Bayesian factor analysis models including MOFA and its spatially structured extension MEFISTO \cite{Argelaguet2018, Argelaguet2020, Velten2022mefisto}. MEFISTO introduces Gaussian process priors that allow latent factors to vary smoothly across spatial coordinates.

Probabilistic and statistical inference methods offer a rigorous and interpretable framework for multimodal spatial omics integration, enabling principled deconvolution, cross-modality alignment, and uncertainty quantification across heterogeneous spatial datasets.

\subsection{Matrix Factorization and Latent Variable Models}

Matrix factorization and latent variable models integrate multimodal spatial omics data by decomposing high-dimensional observations into low-rank representations that capture shared biological structure across modalities. These methods assume that spatial measurements from different assays can be explained by a small number of latent factors corresponding to cell types, gene programs, or spatial patterns, providing an interpretable and computationally efficient framework for integration.

A common formulation is the joint or integrative non-negative matrix factorization, where each modality-specific data matrix $X^{(m)} \in \mathbb{R}^{N \times G_m}$ is approximated as
\begin{equation}
X^{(m)} \approx Z W^{(m)},
\end{equation}
with $Z \in \mathbb{R}^{N \times K}$ representing shared latent factors across spatial locations and $W^{(m)} \in \mathbb{R}^{K \times G_m}$ capturing modality-specific loadings (Figure 4B). Non-negativity constraints encourage additive and biologically interpretable factors. CellPie applies this joint NMF formulation to spatial transcriptomics, enabling spatial domain identification by jointly modelling spatial transcriptomics gene expression and morphological features derived from paired H$\&$E images~\cite{georgaka2025cellpie}. Similarly, LIGER uses integrative NMF approach to jointly factorize multiple unpaired datasets into shared and modality specific factors~\cite{Welch2019}. Latent factor formulations are used in SIMO and SEPAR, which integrate spatial transcriptomics with single-cell or spatial multi-omics data to jointly perform spatial domain identification and, in some cases, deconvolution \cite{yang2025spatial, Zhang2025-mo}. SPOTlight adopts a related NMF-based strategy but incorporates prior knowledge through marker genes from scRNA-seq datasets, yielding seeded latent topics that correspond directly to known cell types \cite{ElosuaBayes2021}.

Several recent methods combine matrix factorization with deep learning components. STRIDE and SpaFusion incorporate NMF-like latent representations together with neural networks to improve robustness and capture nonlinear structure, while still retaining interpretable factor-based decompositions \cite{Sun2022-mi, chen2025spafusion}. COSMOS similarly blends latent variable modeling with deep learning to integrate spatial transcriptomics with spatial proteomics or epigenomics for spatial domain identification and deconvolution \cite{zhou2025cooperative}. 

Matrix factorization and latent variable models provide an interpretable middle ground between purely probabilistic inference and fully deep learning–based approaches. By explicitly modeling shared and modality-specific factors, they are well suited for exploratory multimodal spatial analysis, spatial deconvolution, and identification of dominant biological programs across heterogeneous spatial omics datasets.

\subsection{Optimal Transport and Geometric Alignment}
Optimal transport (OT) and geometric approaches tackle spatial multi-omics integration from a distance and structure preserving perspective. In this paradigm, spatial locations, cells and molecular profiles are represented as empirical measures embedded in a metric space, and cross-modal or cross-sample alignment is formulated as an optimization problem that minimizes transport cost under a prescribed cost function. This geometric view is particularly attractive for spatial omics data, where spatial coordinates, tissue architecture and neighbourhood relations carry as much information as the expression profiles themselves.

\begin{figure}
    \centering
    \includegraphics[width=1.1\linewidth]{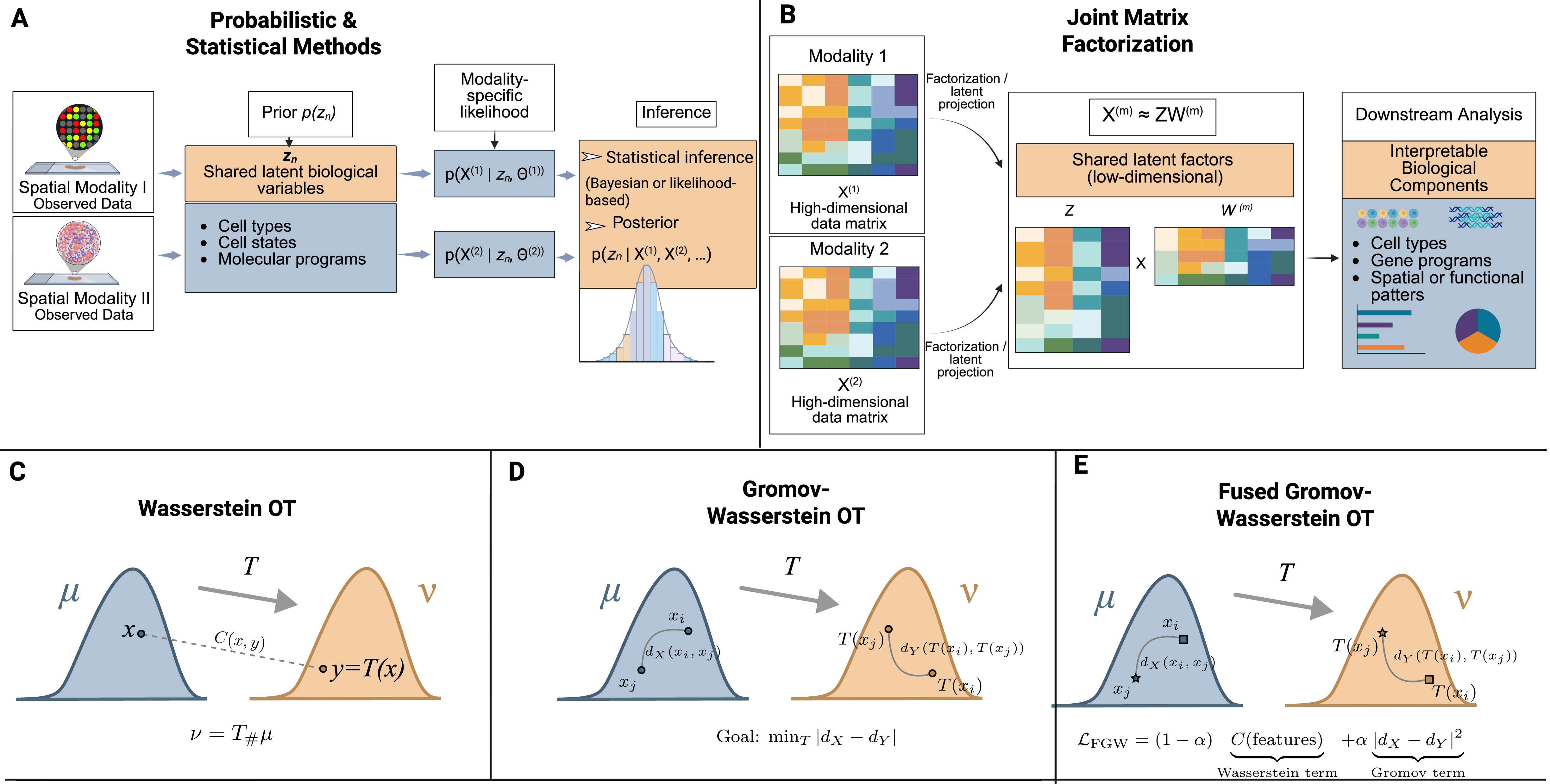}
    \caption{\textbf{Computational frameworks for multimodal spatial omics integration.} \textbf{A:} Probabilistic and statistical inference methods.
    Multimodal spatial omics data are modeled using shared latent biological variables \( z_n \) associated with each spatial location \( n \). These latent variables capture underlying biological structure such as cell types, cell states, or molecular programs. Each modality is generated from \( z_n \) through a modality-specific likelihood \( p(X^{(m)} \mid z_n, \Theta^{(m)}) \), while prior distributions \( p(z_n \mid \Phi) \) encode biological or structural assumptions. Bayesian or likelihood-based inference is used to estimate posterior distributions \( p(z_n \mid X^{(1)}, X^{(2)}, \ldots) \), enabling uncertainty-aware multimodal integration. \textbf{B:} Matrix factorization and latent variable models. High-dimensional data matrices from multiple modalities are decomposed into shared low-dimensional latent factors \( Z \in \mathbb{R}^{N \times K} \) and modality-specific loading matrices \( W^{(m)} \in \mathbb{R}^{K \times G_m} \), such that \( X^{(m)} \approx Z W^{(m)} \). The shared latent factors capture common biological structure across modalities and support interpretable downstream analysis, including identification of cell types, gene programs, and spatial or functional patterns. \textbf{C:} Wasserstein optimal transport (OT) quantifies the minimal cost to transform one distribution $\mu$ into another $\nu$ through a transport map $T$, where the cost is defined by a function $C(x,y)$ measuring the dissimilarity when matching samples from $\mu$ and $\nu$, enabling alignment of datasets in a common feature space. \textbf{D:} Gromov-Wasserstein OT aligns datasets across different feature spaces by preserving pairwise distance relationships within each modality, matching points based on structural similarity rather than direct feature correspondence, with the goal of finding a map $T$ that minimizes $\left| d_X - d_Y \right|$. \textbf{E:} Fused Gromov-Wasserstein OT combines feature-level costs (Wasserstein term) with structure-preserving costs (Gromov term), weighted by parameter $\alpha$, enabling simultaneous alignment of expression similarities and spatial or morphological distances.}
    \label{fig:placeholder}
\end{figure}

Current methods based on Optimal Transport (OT) rely on the Wasserstein distance to quantify differences between distributions in a common feature space, interpreting this difference as the minimal “effort” required to transform one distribution into the other \cite{villani2008optimal}. For example (Figure 4C), OT can relate a set of single cells to a set of spatial spots by estimating a transport plan $\pi$, which can be interpreted as a weight (or probability) matrix linking cells to spots. The plan is computed to minimize a cost that combines transcriptomic dissimilarity and spatial proximity, while ensuring that the total mass assigned to cells and spots remains consistent.




At present, most Wasserstein OT methods are mainly used for in-silico cell-to-space mapping (mapping dissociated single cells to spatial coordinates). For example, SpaOTsc uses structured OT to couple scRNA-seq with spatial data and infer spatially informed cell–cell distances/communication \cite{cang2020inferring}, and novoSpaRc learns a probabilistic cell-to-location assignment for spatial reconstruction \cite{moriel2021novosparc}.

However, many spatial multi-omics tasks require aligning datasets that are represented in distinct feature spaces, such as transcriptomic and proteomic measurements, or gene expression and histology-derived features. Gromov–Wasserstein (GW) OT and its fused variants address this setting by matching the internal relational structure of two datasets rather than their raw coordinates \cite{villani2008optimal}. As illustrated in Figure 4D, instead of directly penalizing distances between individual points, GW minimizes discrepancies between pairwise distance matrices, encouraging correspondences that preserve local similarity structure and underlying topology. Figure 4E further illustrates that Fused GW further incorporates feature-level costs, allowing simultaneous alignment of expression similarities and spatial or morphological distances. 

By encoding spatial coordinates, neighbourhood graphs or anatomical annotations into the cost tensor, GW-based methods can preserve tissue continuity, boundaries and layered organization during integration, and are naturally suited to handling resolution mismatches, multi-section co-registration and cross-modal alignment. However, the current body of GW/FGW-driven spatial integration work remains relatively small, with representative examples including PASTE/PASTE2, which use fused Gromov–Wasserstein OT to align adjacent spatial transcriptomics sections and integrate multiple slices into a consensus (center) slice or support 3D stacking/reconstruction \cite{zeira2022alignment,liu2023paste2}; Moscot, a scalable OT framework that supports FGW-based multimodal profiles and spatial dataset alignment \cite{klein2025mapping}; and SIMO, which takes spatial transcriptomics and multi-omics single-cell datasets as input, and performs FGW-OT–based probabilistic cell–spot alignment for spatial mapping and downstream analyses \cite{yang2025spatial}.

In parallel to OT-based alignment, geometric alignment toolsets integrate spatial multi-omics via explicit cross-section co-registration and resolution harmonisation. For example, Multi-Omics Imaging Integration Toolset (MIIT) first performs histology-guided nonrigid section registration (using GreedyFHist) to align serial sections, then maps ST and MSI onto a common grid and aggregates MSI pixels into each corresponding ST spot weighted by the overlap area, thereby fusing ST and MSI data. This corresponds to a deterministic geometry-driven matching, rather than learning a transport plan by optimizing an OT objective~\cite{wess2025spatial}.

\subsection{Deep Learning and Foundation Models}
Deep learning models now play a central role in multimodal spatial omics integration, providing flexible neural architectures that can capture the heterogeneous, non-linear and spatially structured relationships inherent to these complex and high-dimensional datasets. Deep learning has recently undergone a major paradigm shift with the development of foundation models. Trained on vast, heterogeneous datasets using self-supervised learning, these models acquire general-purpose representations that can be rapidly adapted or fine-tuned to a wide range of downstream tasks \textrm{--}including multimodal spatial omics\textrm{--}using only modest amounts of task-specific data. 
\newline
\textbf{Convolutional Neural Network Models}
\newline
Convolutional Neural Networks (CNNs) have been widely used in multimodal spatial omics integration as versatile encoders of histological images, serving two primary roles: directly as backbone architectures for predicting spatial omics profiles from H$\&$E images, and indirectly as feature extractors that provide morphology-informed inputs to downstream integration architectures. CNNs are hierarchical models typically composed of stacked convolutional layers that progressively extract increasingly abstract features from input images~\cite{CHEN2025109507, lecun2015deep}. Early convolutional layers extract local features (such as texture) through learned filters, while deeper layers combine these early features into more abstract, high level-features, allowing the model to learn more complex tissue structures~\cite{CHEN2025109507, 726791}. Spatial dimensionality of the input images is reduced through pooling operations, improving computational efficiency while retaining important features. The resulting feature maps are transformed into vector representations and passed to a fully connected layer that produces task-specific outputs~\cite{taye2023theoretical, CHEN2025109507, 726791} (Figure 5A). During training, model parameters are learned through gradient-based optimization, with backpropagation used to update the weights~\cite{taye2023theoretical, CHEN2025109507} . 

Among the earliest methods, ST-Net~\cite{he2020integrating} uses CNN-based architecture to learn a mapping between histology image patches and spatial gene expression, enabling the prediction of spatial gene expression profiles from histology. STASCAN predicts spatial cell-type distributions, allowing cell-type composition to be inferred in regions not directly assayed by low-resolution spatial transcriptomics. This enables both spatial resolution enhancement and 3D map construction using imputation on consecutive tissue sections~\cite{wu2024stascan}. At the gene level, analogous CNN-based approaches such as DeepSpaCE predict enhanced spatial gene expression profiles directly from H$\&$E images and enable the imputation of gene expression maps between adjacent tissue sections~\cite{monjo2022efficient}. Building on these predictive frameworks, Zheng et al.~\cite{zheng2023spatial} developed a CNN-based model that integrates paired Glioblastoma histology images, 10x Visium ST data and clinical outcome to predict the spatial distribution of immune cell-types and tumour aggressiveness from unseen H$\&$E images. 

Beyond CNNs, in the field of spatial proteomics, an alternative generative-predictive approach is provided by Ouroboros, a tool based on Generative Adversarial Neural Networks, that enables simultaneous prediction of spatial proteomics profiles from H$\&$E images as well as H$\&$E image generation from spatial protein profiles~\cite{deshpande2024ouroboros}.These methods enable cost-effective spatial molecular profiling as well as in-silico perturbation experiments.

\textbf{Autoencoders $\&$ Contrastive Learning}

\begin{figure*}[t]%
\centering
\includegraphics[width=1.12\linewidth]{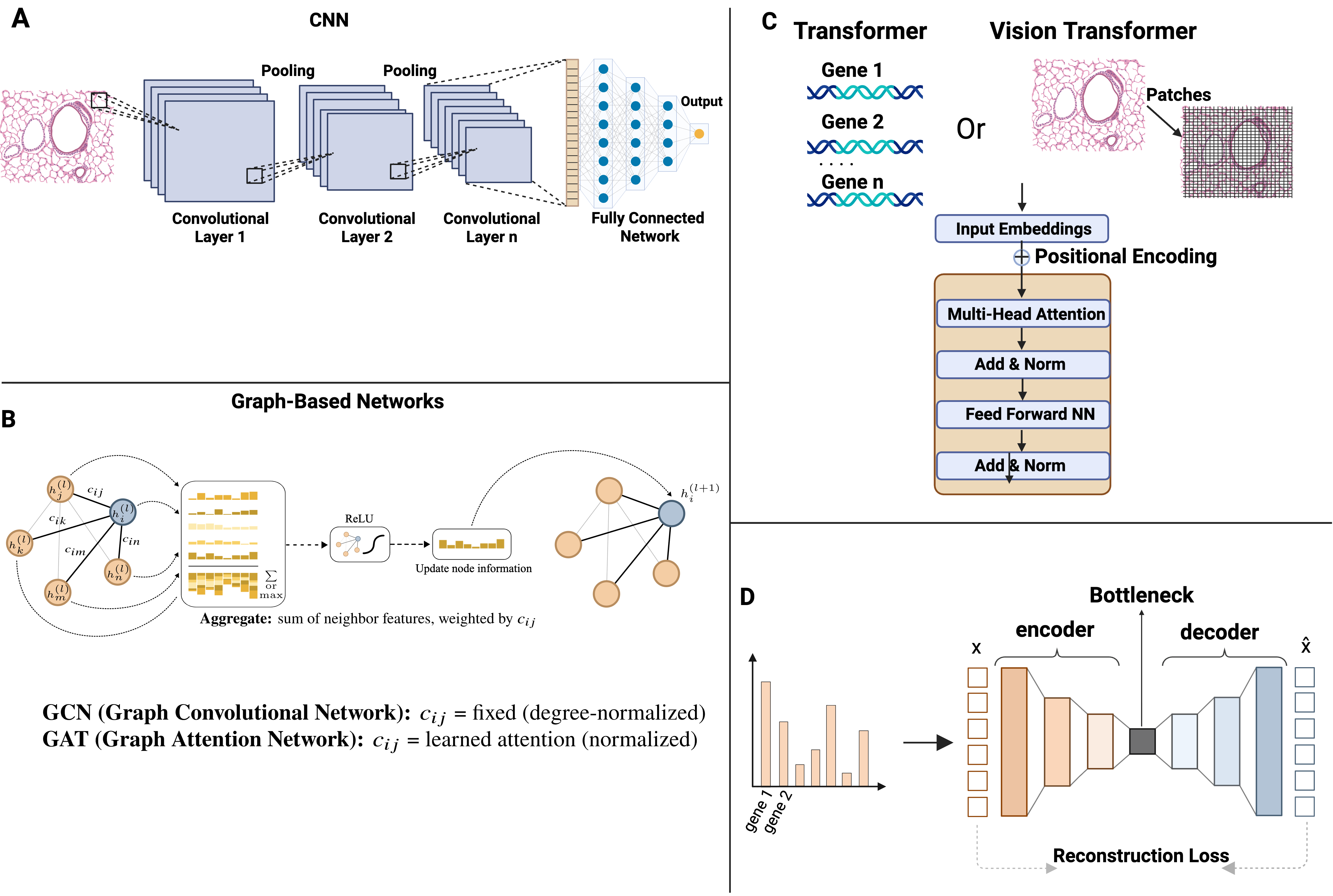}
\caption{\textbf{Deep learning architectures of multimodal spatial omics analysis.} \textbf{A:} CNNs process histology image patches through successive convolutional layers that extract hierarchical features, from low-level textures to high-level tissue structures. Pooling operations reduce spatial dimensionality, and fully connected layers produce task-specific outputs such as gene expression predictions. \textbf{B:} Graph-based neural networks represent tissue locations or cells as nodes with edges encoding spatial adjacency or molecular similarity. In GCNs, node features are updated by aggregating neighbor information weighted by fixed degree-normalized coefficients (c$\textbackslash{}_ij$). GATs expand this by realizing attention weights that adaptively modulate neighbor contributions, enabling the model to focus on the most informative spatial or molecular associations.  \textbf{C:}Transformers use self-attention to capture dependencies over large distances. In spatial omics, genes or proteins can be treated as input sequences, andpositional encodings can maintain spatial information.
Multi-head attention computes pairwise relationships across the sequence, enabling discovery of coordinated gene programs. ViTs adjust this framework for histology images by partitioning images into patches, embedding each as a token, and applying transformer layers to capture global morphological context. \textbf{D:} Autoencoders learn  compact representation of high-dimensional spatial omics data by encoding the inputs into a low-dimensional bottleneck and reconstruct them back through a decoder. The reconstruction loss encourages the latent space to preserve biologically meaningful structure, making autoencoders useful for multi-spatial omic data analysis, data integration, and gene imputation. }
\label{fig:DLM}	
\end{figure*}

Autoencoders (AE) and variational autoencoders (VAE) enable compact representation learning by encoding high-dimensional spatial omics data into a low-dimensional latent space and reconstructing the input through a decoder, as illustrated in Fig. \ref{fig:DLM}D. The learned representations capture biologically meaningful structure because they are optimized via a reconstruction objective that preserves key patterns in the data. VAEs further extend this framework by imposing a probabilistic structure on the latent space through a variational objective (e.g., KL divergence regularization), which encourages smooth, continuous embeddings. 
The power of latent spaces in AEs is well-suited for multimodal integration, as demonstrated in MUSE, where modality-specific embeddings from spatial transcriptomics and images are fused into a shared feature space and refined using reconstruction and self-supervised objectives~\cite{bao2022integrative}. SpaOmicsVAE proposes an architecture to generate a unified representation of any two desired spatial omics data, which are encoded through dual GNNs. It uses the power of VAEs to make the framework more generalizable and to preserve proper distribution within latent space using the Kullback-Leibler divergence loss alongside the reconstruction loss~\cite{zhang2025spaomicsvae}.
In the field of spatial omics, unmeasured gene expression imputation is a challenging question that has been recently addressed using AE architectures. stPlus predicts expression of unmeasured genes by learning the joint embedding between scRNA-seq (reference data) and spatial data while leveraging genes unique to the reference. It then predicts unmeasured gene expression via weighted k-nearest neighbors in the learned latent space ~\cite{shengquan2021stplus}. Along the same line, gimVI introduces a generative VAE model to impute missing genes from unpaired scRNA-seq and spatial transcriptomics by first learning a shared latent cell-state space with modality-specific likelihoods. A shared neural-network decoder maps the latent variable to gene expression proportions. After training, unmeasured genes in spatial cells are imputed by decoding their inferred latent state~\cite{lopez1905joint}. stAI extends the framework of joint-embedding approaches by introducing a dual encoder-decoder architecture that performs both gene imputation and cell-type annotation. stAI first aligns the two datasets by MMD-based latent alignment and improves the imputation using supervised calibration genes, and extra novel loss functions such as gene consistence loss  ~\cite{zou2025stai}. 

In the field of multimodal data analysis, contrastive learning has been an emerging subject of research. Contrastive learning focuses on creating a shared embedding space between paired data modalities by maximizing the similarity between positive pairs, meaning two modalities from the same spot, cluster, or sample, while minimizing the similarity between negative pairs using the cosine similarity loss. Inspired by CLIP~\cite{radford2021learning}, contrastive learning is employed in mclSTExp and BLEEP to learn a joint latent space aligning paired H\&E images and spatial transcriptomics, enabling a new H\&E image to be embedded and its gene expression to be inferred via similarity in the latent space \cite{min2024multimodal, xie2023spatially}. Another approach is to utilize multi-modal contrastive loss function to map H\&E and spatial proteomics (CODEX) measurements into a joint representation space for cell-type annotation~\cite{dayao2025using}. In this setup, positive pairs consist of H\&E-CODEX patches from the same cell, while negative pairs include all other combinations. The contrastive loss encourages embeddings of positive pairs to be close to each other. 
 
\textbf{Graph-based Models}

Within multimodal spatial omics integration, a major branch of deep learning work focuses on graph neural networks (GNNs). In these methods, tissue locations or cells are represented as nodes in a graph, and edges are used to encode spatial adjacency, histology-derived similarity, or neighbourhood relationships at the molecular level. Graph convolutional networks (GCNs), one of the simplest and most widely used GNN architectures, update node representations as illustrated in Figure 5B, by linearly transforming and aggregating the embeddings of neighbouring nodes~\cite{kipf2016semi}. In spatial multi-omics, GCN-based architectures, exemplified by methods such as SpaGCN and GraphST, have been widely used to integrate gene expression, spatial location and histology to identify spatial domains and spatially variable genes~\cite{hu2021spagcn,long2023spatially}. As an extension of GCNs, graph attention networks (GATs) introduce learnable attention weights to adaptively modulate the contributions of different neighbours to node updates~\cite{velivckovic2017graph}. In such models, node updates typically follow the scheme illustrated in Figure \ref{fig:gnn_type}, where the attention coefficients $c_{ij}^{(\ell)}$ are computed from pairwise interactions between neighbouring nodes and can be extended to a multi-head attention mechanism. This enables GATs to focus on the most informative spatial or molecular neighbours and to better characterize heterogeneous microenvironments, sharp tissue boundaries and directional cell–cell interactions. This attention-based design has been widely adopted in spatial multi-omics applications, with methods such as SpatialGlue~\cite{long2024deciphering} and SpatialFuser~\cite{cai2025spatialfuser} exemplifying its use across tasks including identifying spatial domains, expression denoising, and multi-sample integration, highlighting its clear advantages in representing complex spatial microenvironments. In addition to message-passing GNNs, there are also adjacency-graph–centric approaches for multimodal feature extraction and clustering; MISO is one such framework, integrating multiple spatial modalities including spatial transcriptomics, spatial proteomics, spatial epigenomics, spatial metabolomics, and paired histology images~\cite{coleman2025resolving}.

Beyond pre-defined graphs with fixed connectivity, recent work has emphasized learning more adaptive graph structures and modality-specific contributions for paired spatial multi-omics measured on shared coordinates. PRAGA is designed for paired spatial multi-omics integration (such as ST+ATAC or ST+protein) by learning omics-aware graph representations and improving robustness to noisy measurements and unknown annotations, producing unified embeddings for downstream spatial domain discovery and related analyses \cite{huang2025praga}. Similarly, COSMOS targets paired spatial multi-omics integration (e.g., transcriptomics with epigenomics/ATAC or proteomics) by jointly modelling each modality on graphs and combining them with learned modality contributions and contrastive objectives to obtain spatially consistent integrated representations \cite{zhou2025cooperative}. 

Graph models have also been used in cross-modality translation settings, where the goal is to infer an unmeasured spatial modality from an observed one. SpaTranslator performs cross-modal translation across spatial modalities, enabling the prediction of paired spatial multi-omics, notably translating between spatial transcriptomics and spatial epigenomics as well as between spatial transcriptomics and spatial proteomics, by leveraging neighbourhood-aware graph encoders together with generative learning \cite{dong2025spatranslator}. In addition, graph-based designs are increasingly used for imaging–omics bridging; for example, FmH2ST predicts spatial transcriptomics from histological images by combining image-derived features with spatial graph modelling to improve gene expression prediction and denoising \cite{wang2025fmh2st}.

Graph autoencoders are another extension of graph networks, in which they are trained by optimizing the reconstruction loss on the adjacency matrix to provide a way for graph representation learning. SpaFusion utilized the power of this architecture to fuse the graph encodings of spatial transcriptomics and proteomics and perform clustering on the unified representation \cite{chen2025spafusion}.

\textbf{Transformer-based Models}

The transformer architecture, originally introduced for natural language processing (NLP), is based on self-attention mechanisms that enable the modelling of long-range dependencies and the discovery or contextual relationships within sequential data~\cite{vaswani2017attention}, in combination with feed-forward neural networks (Figure 5C). In NLP, transformer-based models operate on a sequence of tokens, where each token usually corresponds to a single word. These tokens are mapped to vector representations, known as embeddings, which encode semantic similarity. As transformer architectures are inherently agnostic to token order, a positional encodings are added to the token embeddings to incorporate positional information of the sequence. Self-attention then computes pairwise similarity scores between these embeddings via scaled dot products, enabling the modelling of contextual relationships across the sequence (Figure 5C). In omics applications, this token-based abstraction can be generalized that each token represents a biologically meaningful unit, for example genes, proteins or metabolites. Applying attention mechanisms to a sequence of omics tokens allows the transformer to learn context-dependent interactions (e.g. gene-gene or protein-protein relationships), facilitating the identification of coordinated biological programs, including gene regulatory networks. Building on this paradigm, spaLLM combines single-cell and spatial omics data by leveraging scGPT, a pretrained single-cell LLM, together with GNNs and multi-view attention aggregation to enhance spatial domain identification \cite{li2025spallm,cui2024scgpt}. Importantly, SpaLLM has been demonstrated across diverse spatial omics technologies, including MISAR-seq, 10x Visium, Spatial-CITE-seq and SPOTS. 

Vision Transformers (ViTs) extend the transformer architecture from sequential data to visual inputs and have recently proposed as an alternative to CNNs \cite{kim2023lightweight, atabansi2023survey, mauricio2023comparing}. Similarly to CNNs, ViTs act as visual encoders of histology images \cite{atabansi2023survey}. In contrast to standard CNNs, ViTs capture global spatial context and long-range dependencies through the self-attention mechanism \cite{yang2022transformers, dosovitskiy2020image, takahashi2024comparison}. Analogous to standard transformers, ViTs treat images as sequences by partitioning them into fixed-size, non-overlapping patches, each of which is then flattened into a 1D vector and converted into a high-dimensional input embedding, while to preserve spatial information, positional encodings are added to each embedding \cite{yang2022transformers, dosovitskiy2020image} (Figure 5C). Global context and long-range dependencies between the image patches are learned through the self-attention mechanisms, followed by a feed-forward network that applies non-linear transformations to each embedding \cite{yang2022transformers, dosovitskiy2020image, takahashi2024comparison}. 

A series of methods incorporating transformers and ViT components have been developed to predict spatial gene expression from histology images. These methods are generally implemented as hybrid architectures combining transformer-based with CNN or Graph-based approaches. TransformerST combines ViTs, Graph-transformers and GCNc to enhance the 10x Visium spatial gene expression resolution to single-cell level, enabling improved clustering performance~\cite{zhao2024innovative}. Similarly, THItoGene adopts a hybrid framework that combines CNNs, ViTs and GATs, while Hist2ST combines CNN, transformer and GNN architecture for spatial gene expression prediction from histology images~\cite{jia2023thitogene, zeng2022spatial}. Related hybrid-based architectures such as MISO leverage ViTs with Local Attention Multiple Instance Learning to predict spatial gene expression from histology~\cite{schmauch2025deep}. Unlike the standard global self-attention, local attention in MISO is restricted to neighbouring instances defined by the k-nearest neighbours based on Euclidean distance between the spatial locations and thus reducing the computational burden. SPATIA is a multi-resolution tool that fuses image-based spatial transcriptomics gene expression with cell-morphology, derived from images, into a unified embedding through cross-attention mechanisms. SPATIA supports both predictive and generative tasks, including cell annotation, gene imputation as well as image generation conditioned on gene expression. In addition, the authors have constructed and released a new large benchmarking dataset for image-based spatial transcriptomics with one-to-one matched morphology and gene expression profiles, spanning multiple donors, tissue and disease types~\cite{kong2025spatia}. 

\textbf{Multimodal Foundation Models}

The development of multimodal foundation models for spatial omics is an emerging field. Recent proof-of-concept studies, include Nicheformer, a transformer-based multimodal foundation model pretrained on massive human and mouse imaged-based spatial transcriptomics and single-cell datasets, comprising over 110 million cells across diverse technologies~\cite{tejada2025nicheformer}. Nicheformer learns a joint representation of spatial and single-cell transcriptomics data through modality-aware contextual tokens, enabling in silico mapping of dissociated single-cell profiles into a spatially informed space. OmiCLIP is a dual-encoder image-transcriptomics foundation model trained on paired 10x Visium spatial transcriptomics data and whole-slide images from $1,007$ tissue samples spanning 32 organs, and supports a range of downstream tasks, including ST gene expression prediction from unseen H$\&$E images, ST and H$\&$E image alignment as well as cell-type deconvolution~\cite{chen2025visual}. OmiCLIP's architecture consists of two modality-specific encoders; an image encoder based on ViT and a transcriptomics encoder based on a causal-masked transformer. Whole-slide histology images are cropped into patches corresponding to the spatial resolution of 10x Visium spots and processed by the image encoder, while for each spot the top 50 expressed genes, ranked by expression level, are converted into a tokenized gene sequence (text) and passed to the transcriptomics encoder (text encoder). The two modalities are then aligned into a shared representation space using a contrastive learning objective.

\section{Challenges and Future Directions}

Despite remarkable progress in spatial multi-omics technologies and computational methods, several fundamental challenges must be addressed before these approaches can achieve their full potential in basic research and clinical translation.

\subsection{Technical Limitations and Standardization}

\textbf{Assay compatibility and tissue preservation}

Simultaneous measurement of multiple molecular layers from the same tissue section remains technically challenging. Sequential processing steps can compromise data quality,for example, immunostaining protocols may degrade RNA integrity for subsequent transcriptomic profiling, while fixation methods optimized for proteomics may not preserve metabolite distributions~\cite{vickovic2022sm, liu2023spatial}. Co-profiling technologies such as DBiT-seq, MISAR-seq, and spatial ATAC--RNA-seq require careful optimization and are not universally compatible across tissue types or preservation methods. The technical complexity of preserving RNA, proteins, chromatin accessibility, and metabolites simultaneously in a single section presents a fundamental biochemical challenge that requires continued protocol development.

\textbf{Resolution heterogeneity}

Different spatial modalities operate at fundamentally different resolutions: array-based transcriptomics comes at a multi-cellular resolution, whereas imaging mass spectrometry achieves or near-single-cell resolution~\cite{angelo2014multiplexed, Giesen2014}, and imaging-based transcriptomics platforms (MERFISH, Xenium) provide subcellular resolution. This resolution mismatch complicates integration, as high-resolution modalities capture cellular and subcellular heterogeneity obscured in lower-resolution measurements. For instance, integrating 2~$\mu$m resolution IMC data with 55~$\mu$m Visium spots requires aggregating the former and deconvolving the latter (Box~1), both of which introduce uncertainties. Emerging approaches---including tissue expansion microscopy \cite{chen2015spatially}, high-density microfluidic barcoding~\cite{vickovic2019high, stickels2021highly}, and highly-resolved ST platforms are narrowing this gap, though complete resolution harmonization across all modalities remains elusive.

\textbf{Modality-specific technical challenges}

Beyond general integration difficulties, each spatial omics modality presents unique technical hurdles. Spatial epigenomics technologies generate inherently sparse data---chromatin accessibility and histone modification measurements exhibit severe zero-inflation even at single-cell resolution. This sparsity is exacerbated in tissue sections, requiring spatial aggregation across neighborhoods to achieve statistical power while potentially masking fine-grained regulatory heterogeneity~\cite{deng2022spatial}. Spatial metabolomics via MSI faces distinct challenges in metabolite identification due to ion suppression effects, matrix interferences, limited spectral reference databases, and the immense chemical diversity of metabolites~\cite{buchberger2018}. Unlike genes or proteins, which have defined sequences, metabolites lack universal barcodes, making confident annotation difficult. Moreover, integrating metabolomics with transcriptomics or proteomics is complicated by indirect correspondence---metabolite levels reflect enzyme activity, transport, and turnover rather than simply gene expression, requiring mechanistic pathway models for interpretation.

\textbf{Cost, throughput, and accessibility}

Multimodal profiling multiplies per-sample costs. Complete tissue characterization combining transcriptomics, proteomics, epigenomics, and metabolomics often exceeds 5,000--10,000 per specimen when accounting for reagents, instrument time, and computational analysis. Specialized instrumentation requirements limit accessibility to well-resourced core facilities and institutions. These capital and operational barriers constrain large-cohort epidemiological studies, longitudinal disease monitoring, and clinical implementation where cost-per-sample budgets are stringent.

\textbf{Lack of standardized protocols}

The field currently lacks consensus protocols for sample preparation, quality control metrics, and data reporting standards. Laboratory-specific workflows hinder cross-study comparisons and reproducibility---one group's ``spatial transcriptomics'' may use fresh-frozen tissue with Visium, while another uses FFPE with GeoMx, yielding fundamentally different data characteristics~\cite{marx2021method}. Batch effects arising from differences in tissue handling, fixation duration, sectioning thickness, staining protocols, and reagent lots are pervasive. Community-driven initiatives to establish best practices, reference materials (e.g., spike-in controls, tissue standards), and metadata annotation guidelines---analogous to efforts by the Human Cell Atlas~\cite{regev2017human} in single-cell genomics---are urgently needed to enable meta-analyses and accelerate method development through rigorous benchmarking. The same applies for H\&E-stained imaging. 

Similarly, H\&E imaging lacks standardized protocols for scanning resolution, color normalization, and image preprocessing. Variations in scanner hardware, objective magnification, compression algorithms, and staining intensity across institutions introduce batch effects that confound downstream image-based predictions and cross-study integration.

\subsection{Computational Complexity and Data Management}

\textbf{Data volume and storage}

Individual spatial multi-omics experiments generate terabyte-scale datasets. For example, a single tissue section profiled by Visium paired with high-resolution H\&E imaging, IMC proteomics, and MALDI metabolomics can easily exceed 1--5$\sim$TB of raw and processed data. Longitudinal studies, cohort analyses, or 3D tissue reconstructions from serial sections amplify these demands. Data storage, transfer, and long-term archiving present non-trivial infrastructure challenges, particularly for academic laboratories without dedicated bioinformatics cores or cloud computing budgets. Standardized, compressed file formats (analogous to BAM for sequencing) and public data repositories with cloud-accessible storage are needed.
 
Beyond infrastructure, the scale of spatial multi-omics data demands computationally scalable methods. Many state-of-the-art approaches---including deep learning architectures and Bayesian inference frameworks---require extensive training data and computational resources that may be prohibitive for smaller laboratories or exploratory studies. Developing lightweight, sample-efficient algorithms alongside data-hungry foundation models remains a critical priority.
 
\textbf{Noise propagation and signal extraction}

Each modality introduces distinct noise characteristics: sequencing dropouts and amplification biases in transcriptomics, antibody cross-reactivity and autofluorescence in imaging, matrix effects and ionization variability in mass spectrometry. When integrating noisy measurements across modalities, there is risk of either amplifying technical artifacts (if modalities share systematic biases) or diluting biological signals (if noise is uncorrelated). Distinguishing true cross-modal relationships from spurious correlations requires robust statistical frameworks that explicitly model modality-specific noise structures~\cite{lopez2018deep}. For example, apparent mRNA--protein discordance may reflect technical noise rather than post-transcriptional regulation, necessitating uncertainty quantification in integration outputs.

\textbf{Analysis pipeline fragmentation}

\paragraph{Analysis pipeline fragmentation}
Current workflows typically involve stitching together disparate software tools: image preprocessing (ImageJ~\cite{schindelin2012fiji}, QuPath~\cite{bankhead2017qupath}), spatial transcriptomics quantification (Space Ranger~\cite{zheng2017massively}, STARmap tools~\cite{wang2018three}), H\&E feature extraction (HistoQC~\cite{janowczyk2019histoqc}), mass spectrometry processing (Cardinal~\cite{bemis2015cardinal}, SCiLS), quality control (custom scripts), and integration methods. This heterogeneity creates steep learning curves, introduces opportunities for error at each handoff, and impedes accessibility for biologists without computational expertise. Development of unified, end-to-end pipelines with standardized inputs/outputs—such as Squidpy \cite{palla2022squidpy} and SpatialData \cite{marconato2024spatialdata}, an open and universal framework for processing spatial omics data—would democratize spatial multi-omics analysis.Cloud-based platforms offering interactive analysis (e.g., Galaxy for spatial data) could further lower barriers.

\subsection{Integration Validation and Benchmarking}

As outlined in Box~1, spatial multi-omics analysis relies on multiple interconnected computational tasks: cell-type deconvolution, spatial domain identification, cross-modality co-registration, histology-based molecular prediction, and gene imputation. However, validating the accuracy of these methods remains a fundamental challenge.

\textbf{Method selection and parameter tuning}

The proliferation of integration algorithms creates a paradox of choice~\cite{argelaguet2021computational}. Methods vary in their assumptions, computational requirements, and suitability for different experimental designs. Clear guidelines for method selection based on data characteristics---sample size, modality types, resolution mismatches, batch structure---are lacking. Practitioners often resort to trial-and-error or follow the most-cited method, which may not be optimal for their specific use case.

\textbf{Absence of ground truth}

Unlike supervised learning tasks with labeled training data, spatial multi-omics integration lacks objective ground truth for validation. When mRNA and protein measurements disagree on cell-type assignment at a spatial location, determining which modality is ``correct''---or whether both capture complementary but valid information (e.g., mRNA in cell body, protein in projections)---is non-trivial. This challenge is particularly acute in diagonal integration, where modalities share no common anchor (e.g., transcriptomics from donor A, proteomics from donor B)~\cite{argelaguet2020mofa}. Without known truth, validation relies on indirect proxies: consistency with orthogonal measurements (immunofluorescence confirming predicted protein), alignment with published cell-type markers, preservation of expected spatial autocorrelation, or biological plausibility of inferred relationships. These heuristics are valuable but cannot definitively prove correctness.

\textbf{Need for systematic benchmarking}

Rigorous benchmarking studies are essential to evaluate integration accuracy and identify failure modes. Ideal benchmarks would include: (1) simulated datasets with known ground truth (e.g., in silico mixtures of single-cell references with defined cell-type compositions and spatial patterns); (2) spike-in controls where synthetic barcodes or proteins are added at known locations; (3) orthogonal validation cohorts profiled by entirely independent technologies; and (4) perturbation experiments where biological ground truth is established (e.g., genetic knockout verified by genomic sequencing)~\cite{luecken2022benchmarking}. Community challenges—similar to DREAM competitions~\cite{saez2016crowdsourcing} or CAFA~\cite{radivojac2013large} in systems biology—could crowdsource method development and establish performance baselines across diverse tissue types and biological questions.

\subsection{Biological Interpretation and Multi-Scale Integration}

\textbf{Cross-layer molecular reasoning}

Interpreting multi-omics patterns requires integrating knowledge across molecular layers with distinct regulatory relationships. Chromatin accessibility enables transcription factor binding; mRNA abundance influences but does not determine protein levels due to translational regulation and protein stability; protein enzymes catalyze metabolite transformations \cite{vogel2012insights}. Deviations from expected layer-to-layer correlations may reflect biologically meaningful regulation (e.g., miRNA-mediated post-transcriptional silencing causing mRNA--protein discordance) or technical artifacts (e.g., antibody cross-reactivity inflating protein measurements). Adjudicating these scenarios demands mechanistic pathway knowledge, time-series data to infer directionality, and perturbation experiments to establish causality---capabilities often unavailable in cross-sectional spatial studies.

\textbf{Multi-scale spatial organization}

Tissues exhibit nested hierarchical structures: molecules within subcellular compartments, compartments within cells, cells within microenvironmental niches, niches within anatomical regions. Observed spatial patterns may arise from multiple non-mutually-exclusive mechanisms: cell-intrinsic transcriptional programs (e.g., zonated hepatocyte metabolism), local cell--cell signaling (paracrine gradients), or tissue-scale physical forces (oxygen/nutrient gradients). Disentangling these influences requires multi-scale statistical models that partition variance across spatial length scales and experimental perturbations that isolate specific mechanisms~\cite{cang2020inferring}. Current methods often operate at a single scale (spot-level or cell-level), missing emergent properties at tissue or organ scales.

\textbf{Data visualization and accessibility}

Effective communication of spatial multi-omics results to diverse audiences---computational biologists, experimental researchers, clinicians, and the public---requires intuitive visualization strategies. Challenges include: (1) dimensionality: conveying information across genes, proteins, metabolites, and spatial coordinates simultaneously; (2) scale: enabling exploration from tissue-level maps down to single-cell detail; (3)interactivity: allowing users to query specific genes/proteins, toggle modalities, and overlay clinical metadata \cite{stuart2019comprehensive}. Interactive tissue atlases~\cite{nitzan2019gene, solorzano2020tissuumaps} with searchable gene/protein databases, linked spatial maps (e.g., Allen Brain Atlas style interfaces) \cite{leinweber2009}, and 3D tissue reconstructions \cite{wang2018three} from serial sections represent the state-of-the-art. However, many published datasets lack accessible visualization portals, limiting their utility for hypothesis generation by the broader community.

\textbf{Toward predictive and mechanistic spatial models}

Realizing the full translational potential of spatial multi-omics requires moving beyond descriptive mapping to predictive modeling. Key aspirational goals include: (1) in silico perturbation: simulating how genetic knockouts, drug treatments, or microenvironmental changes would propagate across molecular layers in spatial context; (2) biomarker discovery: identifying minimal gene/protein signatures that predict clinical outcomes (response to therapy, metastatic potential) from spatial architecture; (3)digital twins: constructing mechanistic models of tissues that recapitulate observed spatial multi-omics data and enable counterfactual queries~\cite{lotfollahi2019scgen}. These applications demand integration of spatial multi-omics with dynamic models (ordinary/partial differential equations, agent-based simulations) and causal inference frameworks---methodological frontiers that remain largely unexplored.

\subsection{Future Outlook}

Despite these challenges, the field is poised for transformative advances. Lessons from single-cell genomics---where analogous challenges in scalability, batch correction, and multimodal integration were systematically overcome through community collaboration, open data sharing, and sustained investment---provide an encouraging roadmap. Key priorities for the spatial multi-omics community include:

\begin{itemize}
\item \textbf{Standardized protocols and reference datasets}: Development of consensus sample preparation workflows, spike-in standards for quality control, and benchmark datasets with ground truth annotations (simulated and experimental) to enable rigorous method comparison.

\item \textbf{Open data repositories}: Expansion of public archives (e.g., spatial extensions to the Human Cell Atlas~\cite{regev2017human}, Human Tumor Atlas Network~\cite{rozenblatt2020human}) with rich metadata annotation (tissue provenance, processing protocols, clinical covariates) and cloud-accessible storage to facilitate meta-analyses and machine learning model training.

\item \textbf{User-friendly analysis platforms}: Creation of integrated software ecosystems (extending Seurat, Scanpy, Giotto, Squidpy) with graphical interfaces, automated quality control, and best-practice pipelines that lower barriers for experimental biologists.

\item \textbf{Clinical translation pathways}: Reduction of per-sample costs through automation, miniaturization, and economies of scale; streamlining of workflows to fit clinical timeframes (diagnosis within days); and regulatory frameworks for spatial omics-based diagnostics.

\item \textbf{Interdisciplinary collaboration}: Bridging technology developers, computational method developers, and disease-focused researchers through consortia, workshops, and collaborative grants to ensure methods address real biological and clinical needs rather than computational benchmarks alone.
\end{itemize}

As experimental resolution improves (approaching true single-cell and subcellular measurements across all modalities), costs decline (through technological maturation and competition), and computational methods mature (validated on benchmarks and applied at scale), spatial multi-omics will transition from a specialized research tool to a routine approach for dissecting tissue biology. The convergence of nanoscale imaging, single-molecule sequencing, machine learning, and mechanistic modeling positions the field to deliver on its promise of revolutionizing our understanding of tissue organization in development, homeostasis, and disease---ultimately enabling precision medicine grounded in the spatial architecture of human tissues.


\bibliographystyle{unsrt}  
\bibliography{getwriting}


\end{document}